\documentclass[showpacs,prb,twocolumn,floatfix,amsmath,,amssymb,superscriptaddress]{revtex4}
\usepackage{graphicx,amsmath,,amssymb}
\begin{document}
\title{Spin properties of a two dimensional electron system: \\ valley 
degeneracy and finite thickness effects}
\author{R. K. Moudgil}\email 
[]{rkmoudgil@kuk.ac.in}  
\affiliation{Department of Physics, Kurukshetra University,
  Kurukshetra - 136 119, India}
\author{Krishan Kumar}
\affiliation{Department of Physics, Kurukshetra University,
  Kurukshetra - 136 119, India}
\affiliation{P.G. Department of Applied Physics, S. D. College,
  Ambala-Cantt. - 133 001, India}
\author{Gaetano Senatore} 
\affiliation{Dipartimento di Fisica, Universit\`a di Trieste, Strada
Costiera 11, 34151 Trieste, Italy}
\affiliation{CNR--IOM DEMOCRITOS Simulation Center,  Trieste, Italy}
\date{\today} 
\begin{abstract} The spin susceptibility of a two-dimensional 
electron system is calculated by determining the spin-polarization 
dependence of the ground-state energy within the
self-consistent mean-field theory of Singwi {\it et al.} (STLS). 
Results are presented for
three different devices, viz. 
the Si (100) inversion layer, the AlAs quantum well, and 
the GaAs heterojunction-insulated gate
field-effect transistor.  
We find a fairly good agreement with experiments for the Si (100) system, on most of the
experimental density range, whereas the agreement for the AlAs and GaAs systems
is less satisfactory;
in all cases, however, it is vital to include the
characteristic device parameters like the   
valley degeneracy, the finite transverse thickness, etc. 
Further, the STLS theory predicts an abrupt
spin-polarization transition at a sufficiently low electron density
irrespective of the valley degeneracy and/or the finite thickness, with 
 the  partially spin-polarized states remaining unstable.
Moreover, in the Si (100) inversion layer, the spin-polarization
transition is
preceded by the simultaneous valley- and spin-polarization;
for its zero thickness model, these transitions however
grossly disagree with the recent quantum Monte Carlo simulations.   
This drawback of the STLS theory is traced to its  inaccuracy
in treating electron correlations, which in turn become more and more
important as the number of independent components (spin and valley) increases.  
\end{abstract}  
\maketitle
\section{Introduction}   
Recently, the study of spin properties of a
two-dimensional electron system (2DES), 
realized at the interface of semiconductor-insulator/semiconductor 
heterojunctions \cite{ando1}, 
has attracted a great deal of interest. 
Envisaged as having its fundamental as well as technological
importance (e.g., in spintronics \cite{awscholom}), one of the chief 
concerns has been to
determine the 2D spin phase diagram. In this regard, 
the recent quantum Monte Carlo (QMC) simulations \cite{sena1,attacc}
have revealed that the electron correlations
may favor a weakly first-order 
spin-polarization transition at $r_s\approx 26$ before the evolution
of Wigner crystallization at
$r_s\approx 35$. 
Unlike in a 3DES \cite{ortiz}, the partially spin-polarized phases  were
found to be unstable.
Here, $r_s=1/(a_B\sqrt{n\pi})$, is the usual coupling
parameter with $n$ as the areal electron density and  
$a_B$ the effective Bohr atomic radius.  
Experimentally, the spin properties have been probed mainly by 
measuring the spin susceptibility $\chi_s$. 
Quite generally, $\chi_s$ has been found to grow over
its Pauli value $\chi_P$ with increasing $r_s$, the growth
depending markedly on the device hosting the 2DES. 
For the Si (100) inversion layer, Shashkin {\it et al.}
\cite{shas1,shas2} and Vitkalov {\it et al.}  \cite{vitka} have
found 
an indication of a ferromagnetic instability at a density close
to the apparent metal-insulator transition (MIT), thus suggesting
a relation between the spin-polarization and the MIT.
However, Pudalov {\it et al.} \cite{puda1} have ruled out
the  possibility of such an instability for densities down to
the MIT, though admittedly in different samples. 
A qualitatively similar result has been found 
for  2D electrons in the  AlAs/GaAs based 
heterostructures  \cite{vakili,zhu}.
\par 
The role of the valley degree of freedom $g_v$
has also been explored \cite{shkol} and, contrary to common expectation 
(based on Hartree-Fock), the spin susceptibility enhancement
$\chi_s/\chi_P$ in the two-valley (2V) state is found to be
smaller than the one-valley (1V) result. Note however that, at least at 
high density the spin susceptibility  $\chi_s$ of the 2V system is indeed larger 
than the 1V one.
 Apart from this, the recent QMC simulations  of the 2V2DES
 due to Marchi  {\it et al.} \cite{marchi} 
 have  shown that the spin properties undergo a qualitative 
change with $g_v$. Notably, the 2V2DES 
did not support the spin-polarization transition, a finding at
variance with the 1V QMC result \cite{attacc}. The discrepancy 
with the experimental  indication of a diverging spin susceptibility
 found by  Shashkin {\it et al.} \cite{shas2}, on the contrary, is 
naturally ascribed to absence of disorder in the QMC simulations.
 In Ref. \onlinecite{marchi}, an analytical fit to the QMC correlation 
energy at arbitrary spin polarization $\zeta$ 
has also been provided; 
$\zeta=(n_{\uparrow}-n_{\downarrow})/(n_{\uparrow}+n_{\downarrow})$,
with $n_{\sigma}$ being the density
of electrons having spin $\sigma$.
\par
Since the measurement of $\chi_s$, there have been theoretical 
attempts \cite{yarla,depalo,zhang} to elucidate the experimental data.  The QMC
simulations \cite{attacc,marchi} have  so far been performed only for 
an ideal (i.e., zero
transverse thickness) 2DES. However, the device-specific parameters
such as the transverse thickness of the electron layer, valley
degeneracy etc., have been shown \cite{depalo,zhang} to have an 
appreciable effect on $\chi_s$.   
Theoretically, the main problem has consisted in treating the electron
correlations, thus making the use of approximations inevitable. 
\mbox{Rajagopal {\it et al.}\cite{rajagopal}} 
first determined the spin properties beyond 
the Hartree-Fock  approximation (HFA) by
numerically evaluating the sum of ring diagrams. 
Zhang and Das Sarma \cite{zhang}
recently used the random-phase approximation (RPA) to deal with
the correlations. The correlations were found to
introduce quantitative as well as  qualitative changes in the 
ground state. Particularly, the critical $r_s$ for transition to the
ferromagnetic phase increased both for the 1V and 2V states, and the 
2V $\chi_s/\chi_P$ now lays
below the 1V result as in simulations \cite{marchi} and
experiment\cite{shas2}. 
Davoudi and Tosi \cite{davoudi} went beyond the RPA by including the
short-range correlations within the 
self-consistent mean-field theory of Singwi, Tosi, Land, and
Sj\"{o}lander (STLS)\cite{stls}, but they  restricted their study  
to an ideal 1V2DES model. 
The ferromagnetic instability was found to occur at almost the same $r_s$
($\sim 5.6$) as in the RPA. However,  
the ground-state energy for the partially spin-polarized
states could be computed only up to $r_s\approx 4$ due to the
appearance of an unphysical instability and accordingly, the fate 
of these states was not clear for $r_s>4$.      
Although no direct comparison with the RPA was
drawn, yet it is known that the RPA is accurate only at low $r_s$. 
\par
A theoretical study of the spin properties of a 2DES  by taking
into account the correlations and the device details (the transverse
thickness and the valley degeneracy) makes the main aim of the present work.  
The correlations will be
treated at the level of STLS approximation.
The STLS has been used previously to study these effects
\cite{jonson,gold1},  but only in the unpolarized phase.    
Firstly, we intend to examine the ground state for its stability with
respect  to $\zeta$ and $g_v$
by comparing the ground-state energy in different states.  
Secondly, we shall extract $\chi_s$ from the $\zeta$ dependence of 
ground-state energy. To assess the role of device details, 
$\chi_s$ will be calculated for three different 
experimentally studied devices, namely,
the Si (100) inversion layer \cite{shas2,puda2}, the AlAs quantum well
(QW) \cite{vakili}, and the GaAs
heterojunction-insulated gate field-effect transistor (HIGFET) \cite{zhu}. 
For the ideal 2DES model, we shall
compare our results with the recent QMC simulations\cite{attacc,marchi}. 
To the best of
our knowledge, the STLS theory has not been so far tested against the
QMC correlation energy at arbitrary $\zeta$. 
\par 
The rest of the paper is organized as follows: The 2DES
model and  the STLS formalism are presented in Sec. II. 
Results and discussion are
given in Sec. III. In Sec. IV the paper is concluded with a brief summary.   
\section{Theoretical formalism}
\subsection{2DES model}
The electrons confined dynamically to a plane,  immersed 
in a rigid charge neutralizing background, with  
$e^2/r$ interaction has been the commonly used
model to study a 2DES. However, the electrons in actual devices 
are trapped in a finite QW along the transverse direction, which
results in a  multi-subband 
 2DES having a finite transverse thickness. 
Quite often,  it is reasonable
to assume that the electrons occupy only the lowest energy subband.
Under these conditions, one has to deal with an effective electron
interaction potential \cite{ando1}
\begin{equation} 
V(q)=\frac{2\pi e^2}{\epsilon q}F(q),
\label{e4} 
\end{equation}
where $\epsilon$ is the dielectric constant of the substrate 
and $F(q)$ the form factor arising due to the finite 
thickness of the 2DES. 
Such a 2D model is often referred to as a quasi 2DES.
On setting  $F(q)=1$, the ideal 2DES model is recovered. 
Obviously, $F(q)$ would depend upon the details of the device hosting
the 2DES. For the Si (100) inversion layer, the system of main interest in the
present study, we use the model of Stern and Howard \cite{stern} where
\begin{eqnarray}
 F(q)&&=\frac{1}{16}\left
(1+\frac{\epsilon_{ins}}{\epsilon_{sc}}\right ) \left (1+\frac{q}{b}\right
)^{-3}\left [8+9\frac{q}{b}+3 \frac{q^2}{b^2}\right ] \nonumber \\
&& +\frac{1}{2}\left (1-\frac{\epsilon_{ins}}{\epsilon_{sc}}\right ) \left
(1+\frac{q}{b}\right )^{-6}, 
\label{e5} 
\end{eqnarray} 
and $\epsilon=(\epsilon_{sc}+\epsilon_{ins})/2$.
Here, $\epsilon_{sc}$ and $\epsilon_{ins}$ are, respectively, the
dielectric constants of the Si substrate and the insulating SiO$_2$
layer, and $b$ is  given by
\begin{equation}
b=\left [\frac{48\pi e^2m_z}{{\epsilon}_{sc}\hbar^2}\left
(n_{d}+\frac{11}{32}n\right ) \right ]^{1/3},
\label{e3}
\end{equation}
with $m_z$ being the electron effective mass in the transverse direction
and $n_{d}$ the electron concentration in the depletion layer.
  Depending upon
the orientation of the Si surface, its conduction band is $g_v$-fold 
degenerate- the so-called valley degeneracy; 
$g_v=2$ for the Si (100) inversion layer.
Thus, if $n_{i\sigma}$ denotes the density
of electrons of spin $\sigma$ in the $i$th valley, we must have:
$n=\sum_{i\sigma}n_{i\sigma}$. Through out our study, we shall take 
$T=0K$.
\subsection{Density response function}
We use the dielectric approach where the density response of the
electron system  to an external electric potential $V^{ext}(q,\omega )$ 
is a central quantity
in determining its ground-state properties. 
It is assumed \cite{gold1,jonson, bloss}  that the
inter-valley scattering is negligibly small.  
The multi-valley electron system 
can then be treated as a multi-component system, with 
the valley index $i$ representing a particular component.
At arbitrary $\zeta$,
the 2V2DES is therefore equivalent to a four-component
system and its linear density response function 
$\chi_{i\sigma i'\sigma'}(q,\omega)$ is defined as
\begin{equation}
\rho^{ind}_{i\sigma}(q,\omega)=\sum_{i'\sigma'}
\chi_{i\sigma i'\sigma'}(q,\omega)V_{i'\sigma'}^{ext}(q,\omega),
\label{e6}
\end{equation} 
where $\rho^{ind}_{i\sigma}(q,\omega)$  represents the induced  
electron density in the 
$i$th valley for the spin component $\sigma$. 
For the response function calculation, we employ the STLS 
mean-field approximation, which has earlier been 
used \cite{sjolander,davoudi} to study 
a two-component system. 
Its extension to the four-component system is 
straightforward and therefore, only the central relations of the 
formulation are given here. 
\par
The induced density
$\rho^{ind}_{i\sigma}(q,\omega)$ can be compactly expressed as 
\begin{equation}
\rho^{ind}_{i\sigma}(q,\omega)=\chi_{i\sigma}^0(q,\omega)\left 
[V_{i\sigma}^{ext}(q,\omega)+V_{i\sigma}^{ind}(q,\omega)\right],
\label{e6a}
\end{equation}
where $\chi_{i\sigma}^0(q,\omega)$ is the
density response function of non-interacting electrons 
having spin $\sigma$ in the $i$th valley 
and 
\begin{equation}
V_{i\sigma}^{ind}(q,\omega)=
\sum_{i'\sigma'} \rho^{ind}_{i'\sigma'}(q,\omega)V(q)
\left [1-G_{i\sigma i'\sigma'}(q) \right ],
\label{e7}
\end{equation} 
is the induced potential. Here,  $G_{i\sigma i'\sigma'}(q)$ are the spin- and
valley-resolved local-field correction (LFC) factors which account for  
correlation among electrons  in valleys $i$ and $i'$ with spins
$\sigma$ and $\sigma'$. In the STLS approach, $G_{i\sigma
  i'\sigma'}(q)$ are given in terms of the corresponding static density
structure factors $S_{i\sigma i' \sigma'}(q)$ as 
\begin{eqnarray}
G_{i\sigma i'\sigma'}(q)= && -\frac{1}{\sqrt{n_{i\sigma}n_{i'\sigma'}}}
\int\frac{d{\bf q'}}{(2\pi)^2}\frac{{\bf q}.{\bf q'}}{q^2}
\frac{V(q')}{V(q)} \nonumber \\
&& \times [S_{i\sigma i'\sigma'}(|{\bf q}-{\bf q'}|)-\delta_{ii'}
\delta_{\sigma\sigma'}].
\label{e8}
\end{eqnarray}
Using Eqs. (\ref{e6})-(\ref{e7}),  
the density response matrix $\chi$ (\mbox{$2g_v\times 2g_v$}) is obtained as
\begin{equation}
\chi={\cal A}^{-1};\, {\cal A}_{i\sigma i'\sigma'}=
 \frac{\delta_{ii'}\delta_{\sigma\sigma'}}{\chi_{i\sigma}^0(q,\omega)}-
V(q) \left[1-G_{i\sigma i'\sigma'}(q)\right].
\end{equation} 
The fluctuation-dissipation theorem \cite{mahan}, which
relates  $S_{i\sigma i' \sigma'}(q)$ with  
$\chi_{i\sigma i'\sigma'}(q,\omega)$ as
\begin{equation}
S_{i\sigma i'\sigma'}(q)=
-\frac{\hbar}{\pi\sqrt{n_{i\sigma}n_{i'\sigma'}}}
\int_{0}^{\infty} d\omega \,\chi_{i\sigma i'\sigma'}(q,\iota\omega),
\label{e10}
\end{equation}
closes the set of STLS equations for the density response matrix. The $\omega$
integration in Eq. (\ref{e10}) has been performed along the 
imaginary $\omega$ axis to 
avoid the problem of plasmon poles on the real  $\omega$ axis \cite{landau}.
Evidently, the response function calculation has to be carried out 
numerically by solving the set of Eqs. (\ref{e8})-(\ref{e10}) in a 
self-consistent way.
\subsection{Ground-state energy and spin susceptibility} 
The ground-state energy  per particle $E_{0}$ can be determined by a straightforward
extension of the ground-state energy theorem \cite{mahan} to the
multi-component system as
\begin{equation}
E_{0}=\frac{1}{n}\sum_{i\sigma}n_{i\sigma}\left
(\frac{\hbar^2k_{F,i\sigma}^2}{4m_b}\right)+\int_{0}^{e^2}\frac{d\lambda}
{\lambda}E^{int}(\lambda),
\label{e11}
\end{equation}
where the first term 
is the kinetic energy per electron of the non-interacting system and
the second term represents the
potential energy  contribution. Here, $m_b$ is the effective band mass,
$k_{F,i\sigma}=(4\pi n_{i\sigma})^{1/2}$, is the Fermi wave vector of
the valley-spin component $i\sigma$, and   
\begin{equation}
E^{int}(\lambda)=\frac{1}{2n} \sum_{i\sigma i'\sigma'}E^{int}_{i\sigma i'\sigma'}(\lambda),
\label{e13}
\end{equation}
is the interaction energy per electron with Coulomb coupling  
$\lambda$. Its components $E^{int}_{i\sigma i'\sigma'}(\lambda)$ are given by 
\begin{eqnarray}
E^{int}_{i\sigma i'\sigma'}(\lambda) &&=
\sqrt{n_{i\sigma}n_{i'\sigma'}}  \int\frac{d{\bf q}}{(2\pi)^2}
 V(q,\lambda)\nonumber \\
&&\times\left[S_{i\sigma i'\sigma'}(q,\lambda)-\delta_{ii'}
\delta_{\sigma\sigma'}\right].
\label{e14}
\end{eqnarray}
Thus, the calculation of $E_0$  is based solely
on the set of static structure factors 
$S_{i\sigma i'\sigma'}(q,\lambda)$.
\par
The  spin susceptibility $\chi_s$ can be obtained from
the second-order derivative of $E_0$ with respect to
$\zeta$ as  
\begin{equation}
\frac{\chi_s}{\chi_P}=\frac{\pi \hbar^2 n}{g_vm_b}\left[\left
(\frac{\partial^2E_0}{\partial \zeta^2} \right )_{\zeta=0}\right ]^{-1}.
\label{e15} 
\end{equation}
Here, $\chi_P=(g_vm_bg^2\mu_B^2/4\pi \hbar^2)$, is
the Pauli susceptibility, with 
$g$ the Lande factor and  $\mu_B$ the Bohr magneton. 
\par
The theoretical formalism given above is applicable to an arbitrary
$g_v$ and $\zeta$. In the next section, we present numerical results 
for $g_v=2$ and $1$.
\section{Results and discussion}
The numerical calculation proceeds as follows: 
Eqs. (\ref{e8})-(\ref{e10}) are solved iteratively  
for the set of
independent partial structure factors $S_{i\sigma i'\sigma'}(q)$. 
These results are then used in 
Eq. (\ref{e11}) to compute the ground-state energy $E_0$ and hence,
the spin susceptibility $\chi_s$ from Eq. (\ref{e15}). 
Except in section III D, the prefix `quasi' with
2DES refers to the Si (100) inversion layer, where the input parameters
used are: $m_z=0.98m_e$, $\epsilon_{sc}=11.8$, $\epsilon_{ins}=3.8$,
and $n_{d}=1.2\times 10^{11}cm^{-2}$.
When two valleys have unequal electron population and partial spin 
polarization, the
calculation of $E_0$ requires ten independent partial $S(q)$'s. 
However, we restrict here to the symmetric case in which  the
two valleys have equal population and 
spin polarization (i.e., $n_{1\uparrow}=n_{2\uparrow}$ and 
$n_{1\downarrow}=n_{2\downarrow}$). With this restriction, the number
of independent partial $S(q)$'s reduce to five, and the computation of  
$E_0$ is greatly simplified as it involves only three independent
combinations of $S_{i\sigma i'\sigma'}(q)$, viz. 
$S_{\uparrow\uparrow}(q)=S_{1\uparrow
  1\uparrow}(q)+ S_{1\uparrow 2\uparrow}(q)$,
$S_{\downarrow\downarrow}(q)=S_{1\downarrow 1\downarrow}(q)+
S_{1\downarrow 2\downarrow}(q)$, and $S_{\uparrow\downarrow}(q)=
2S_{1\uparrow 1\downarrow}(q)$. It turns out that 
$S_{\uparrow\uparrow}(q)$, $S_{\downarrow\downarrow}(q)$, and 
$S_{\uparrow\downarrow}(q)$ are simply the respective $S(q)$'s
of a partially polarized electron system with $g_v$-fold
degenerate $\uparrow$ and $\downarrow$ spin states. 
We begin by discussing $S_{\sigma\sigma'}(q)$, as
$E_0$ is solely determined in terms of these.
Through out the results presented, the wave vector $q$ is in units of 
$k_{F\uparrow}$ ($=k_{F,1\uparrow}=k_{F,2\uparrow} $), energies
in 
effective Rydberg, and $\hbar=1$. 
\subsection{Spin-resolved static structure factors}\label{cf}
We accepted the iterative solution when the convergence in 
$S_{\sigma \sigma'}(q)$ was better than
$10^{-6}$. However, 
it became almost impossible to obtain the self-consistent
solution at and above a critical $r_s$ (say $r_s^c$);
for instance, $r_s^c\approx 3.1$ at $\zeta=0$ for the  quasi 2V2DES. 
We find out that this  
problem arises due to the emergence of (unphysical) poles in
$\chi_{\sigma \sigma'}(q,\iota\omega)$ on the imaginary $\omega$ axis  
over a definite $q$ interval,  $0 \le q \le q_c$. 
A qualitatively similar situation was earlier faced for an ideal
1V2DES by Moudgil {\it et al.} \cite{moud} and  Davoudi
and Tosi \cite{davoudi},  and for a 3DES 
by two of the present authors \cite{kk}. 
Due entirely to these poles,  the authors in
Ref. \onlinecite{davoudi} could compute $E_0$ for the partially 
spin-polarized states only up to $r_s\sim 4$.
However, we have shown in Ref. \onlinecite{kk} 
 that (i) 
the self-consistent $S_{\sigma\sigma'}(q)$ can be obtained for $r_s$ beyond $r_s^c$
if the existence of the poles is taken into account in establishing the
relation between the frequency integrals of 
 $\chi''_{\sigma \sigma'}(q,\omega)$  and $\chi_{\sigma \sigma'}(q,\iota\omega)$, 
(ii)  although the STLS theory
breaks down in describing  $S_{\sigma\sigma'}(q)$ for
$r_s>r_s^c$, yet it provides a fairly good account of the 
total charge-charge structure factor $S_{CC}(q)$, which in terms of
its components is given by
\begin{equation}
S_{CC}(q)=\frac{1+\zeta}{2}S_{\uparrow\uparrow}(q)+ \frac{1-\zeta}{2}
S_{\downarrow\downarrow}(q)+\sqrt{1-\zeta^2}S_{\uparrow\downarrow}(q).
\label{e17}
\end{equation}
It may be pointed out that the calculation of $E_0$ relies solely on the
knowledge of $S_{CC}(q)$ [see Eq. (\ref{e16})]. Taking into account
the likelihood of a pole in
$\chi_{\sigma\sigma'}(q,\iota\omega)$ [say at
\mbox{$\omega=\omega_0(q)$}], 
Eq. (\ref{e10}) takes the form \cite{vin} 
\begin{eqnarray} 
S_{\sigma \sigma'}(q)= && -\frac{1}{\pi\sqrt{n_{\sigma}n_{\sigma'}}} 
\int_{0}^{\infty} d\omega \Big [\chi_{\sigma \sigma'}(q,\iota\omega)\nonumber\\
&&- \frac{a_{\sigma \sigma'}}{\omega-\omega_0(q)} 
+\frac{a_{\sigma \sigma'}}{\omega+\omega_0(q)} \Big ]. 
\label{e18}
\end{eqnarray}  
Here, $a_{\sigma \sigma'}$ is related to the first-order residue  $b_{\sigma
  \sigma'}$ of $\chi_{\sigma \sigma'}(q,z)$ at $z=\iota\omega_0(q)$, by
$a_{\sigma \sigma'}=-\iota b_{\sigma \sigma'}$. 
Making use of the Eq.
(\ref{e18}), we are able to determine the
self-consistent $S_{\sigma\sigma'}(q)$ 
at  any desired $r_s$ and $\zeta$ values.
\par
\begin{figure} [b!]
\includegraphics[width=65mm,height=60mm]{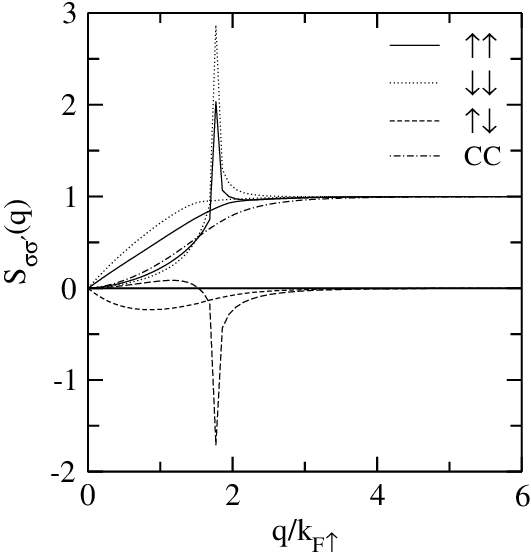}
\caption{\label{fig1} The spin-resolved static structure factors
 $S_{\sigma\sigma'}(q)$ vs $q/k_{F\uparrow}$ at
$\zeta=0.25$, and $r_s=2$ and $4$ (the curves with sharp peaks) 
for the quasi 2V2DES. The dashed-dot curve represents the
total charge-charge static structure factor $S_{CC}(q)$ at
$r_s=4$.} 
\end{figure}
\begin{figure} [t!]
\includegraphics[width=65mm,height=60mm]{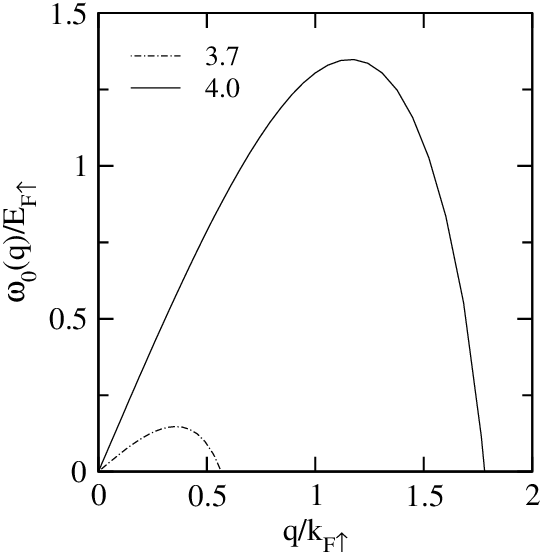}
\caption{\label{fig2} The pole position $\omega_0(q)$ vs $q/k_{F\uparrow}$
at  $\zeta=0.25$ for the quasi 2V2DES.
Legends indicate  the value of $r_s$ and  $E_{F\uparrow}$ is the 
Fermi energy of the $\uparrow$ spin electrons.} 
\end{figure}
\begin{figure} [t!]
\includegraphics[width=65mm,height=60mm]{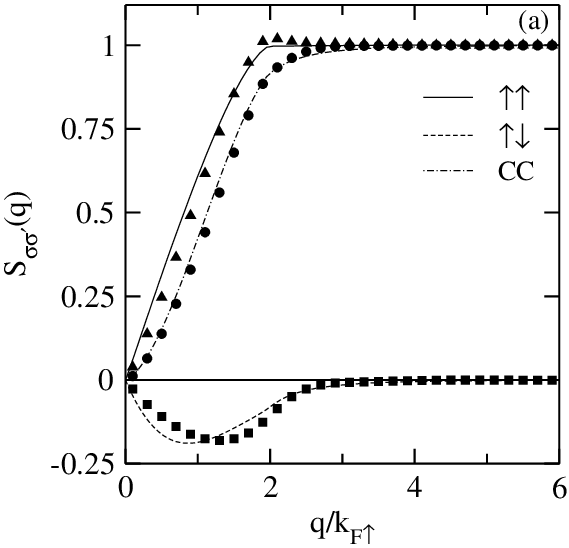}
\includegraphics[width=65mm,height=60mm]{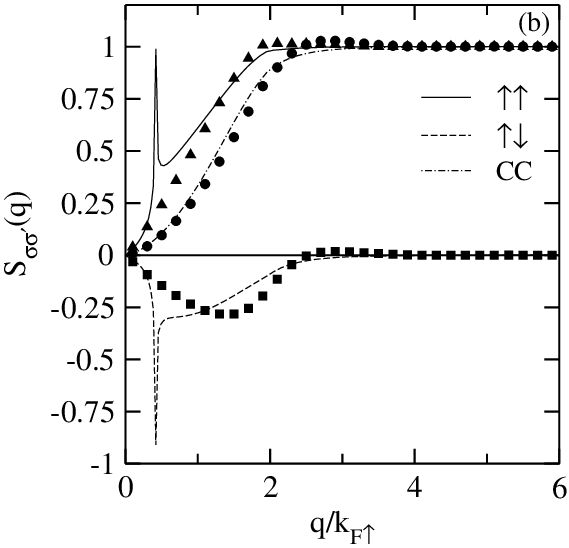}
\caption{\label{fig3} The spin-resolved ($S_{\sigma\sigma'}(q)$) and
the total charge-charge ($S_{CC}(q)$) structure factors
 vs $q/k_{F\uparrow}$ at
$\zeta=0$, and $r_s=2$ [in panel (a)] and $5$ [in panel (b)]  for an 
ideal 1V2DES. The symbols are the QMC data  taken 
from the Ref. \onlinecite{pao}.} 
\end{figure}
We show in Fig. \ref{fig1} the self-consistent
$S_{\sigma\sigma'}(q)$  for the  quasi 2V2DES at $\zeta=0.25$ 
for $r_s=2$ and $4$.   The sharp peaks  
at $r_s=4$ are the artifact of unphysical poles in
$\chi_{\sigma\sigma'}(q,\iota\omega)$.
In order to elaborate this connection, we have
plotted  $\omega_0(q)$ in Fig. \ref{fig2} at $\zeta=0.25$ for 
two $r_s$ values. 
\begin{figure} [t!]
\includegraphics[width=65mm,height=60mm]{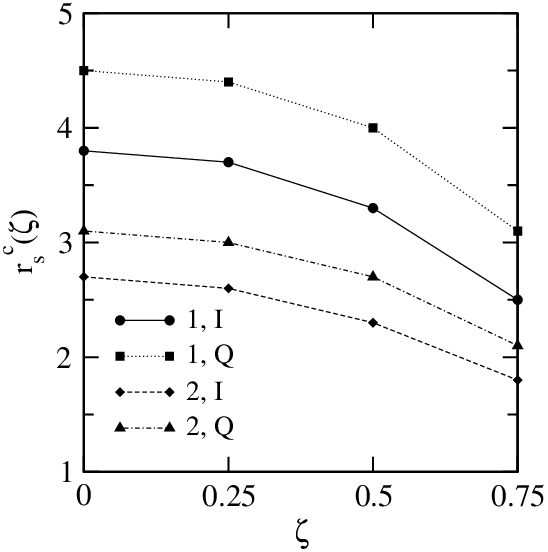}
\caption{\label{fig4} Critical $r_s$ (i.e., $r^c_s$) for the onset of pole in
$\chi_{\sigma\sigma'}(q,\iota\omega)$ vs $\zeta$; in each case the first
legend stands for the value of $g_v$, while the second for the quasi
(Q) or ideal (I) 2DES models. Lines are just a guide to the eye.} 
\end{figure}
Apparently, $S_{\sigma\sigma'}(q)$ have peaks precisely at $q=q_c$ and 
$q_c$ increases with $r_s$.   
 In fact, $S_{\sigma\sigma'}(q)$ start 
exhibiting such peaks for $r_s$ just above $r^c_s$ and 
as $r_s\rightarrow r_s^c$,  both $\omega_0(q)$ and  $q_c$ tend to zero.
Nevertheless, it is interesting to note that
the  resulting $S_{CC}(q)$
remains a smooth function of $q$, implying therefore 
a perfect cancellation of peaks in 
$S_{\sigma\sigma'}(q)$. 
\begin{figure}[t!]
\includegraphics[width=65mm,height=60mm]{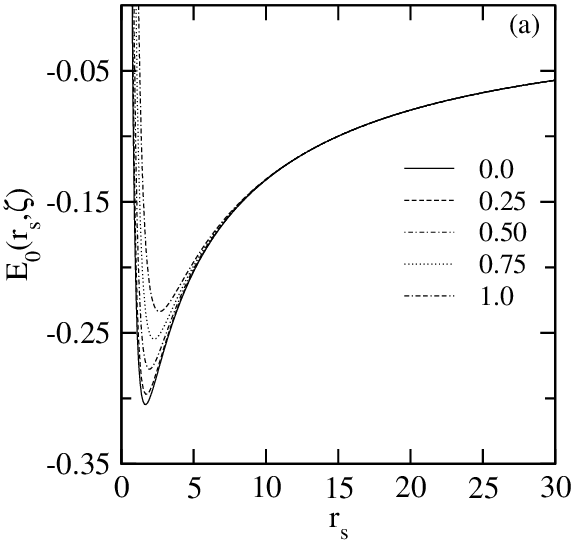}\vspace{5mm}
\includegraphics[width=65mm,height=60mm]{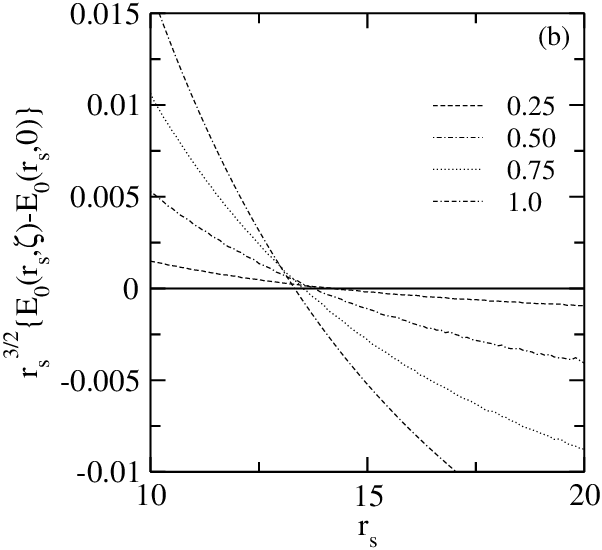}   \caption{\label{fig5}
The ground-state energy $E_0(r_s,\zeta)$ [in panel (a)] and 
$r_s^{3/2}\{E_0(r_s,\zeta )-E_0(r_s,0)\}$  [in panel (b)] plotted
 as a function of $r_s$ at different $\zeta$ for the quasi 2V2DES; legends 
indicate the value of $\zeta$.}  
\end{figure} 
\begin{figure}[b!]
\includegraphics[width=65mm,height=60mm]{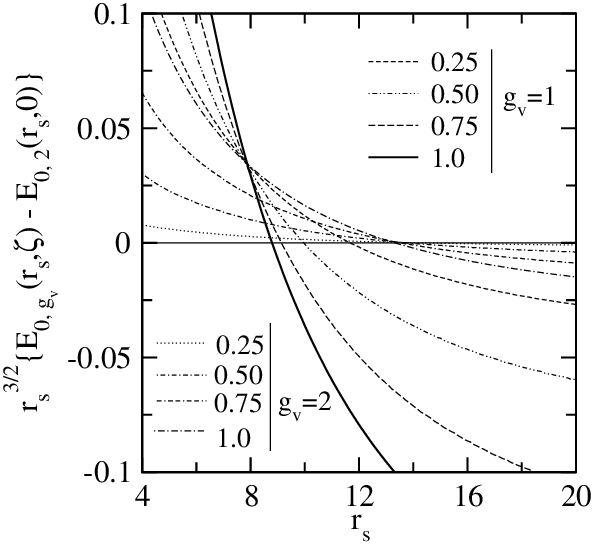} \caption{\label{fig6}
$r_s^{3/2}\{E_{0,g_v}(r_s,\zeta )-E_{0,2}(r_s,0)\}$ plotted
as a function of $r_s$ at different $\zeta$ by taking  $g_v=2$ and $1$
for the quasi 2DES; legends indicate the value of $\zeta$.} 
\end{figure}
We also examine $S_{\sigma\sigma'}(q)$ 
 for an ideal 1V2DES at $\zeta=0$, where QMC simulations \cite{pao} are 
available to check the accuracy  of theory. A direct comparison 
(see Fig. \ref{fig3}) reveals that
both $S_{\sigma\sigma'}(q)$ and $S_{CC}(q)$ agree nicely with 
the QMC data for $r_s< r_s^c$, but thereafter   
only $S_{CC}(q)$ remains close to the simulation
results. Particularly, the sharp peaks in the STLS
$S_{\sigma\sigma'}(q)$ for $r_s> r_s^c$ are in complete contrast with
the QMC study. Thus, as in a 3DES, 
the STLS theory fails to handle the spin-resolved correlations in 2DES
for $r_s > r^c_s$, but it provides a fairly good description of 
the spin-summed correlations. 
 A qualitatively similar picture  is found for
an ideal 2V2DES and the quasi 1V2DES. 
For a ready reference, $r^c_s$  
is given in Fig. \ref{fig4} at selected $\zeta$ for 
the quasi- and ideal-2DESs.
\subsection{Ground-state energy and correlation energy} 
The Eqs. (\ref{e11})-(\ref{e14}) can be combined to 
rewrite  the ground-state energy in reduced units as 
\begin{eqnarray}
E_0(r_s,\zeta) &&= \frac{1+\zeta^2}{g_vr_s^2}+\frac{2}{r_s}
\sqrt{\frac{1+\zeta}{2g_v}} \int_{0}^{1}d \lambda
\int_{0}^{\infty}dq\,F(q) \nonumber \\&& 
\times \left[S_{CC}(q;r_s,\lambda)-1 
\right ]. 
\label{e16} 
\end{eqnarray} 
\begin{figure}[t!]
\includegraphics[width=65mm,height=60mm]{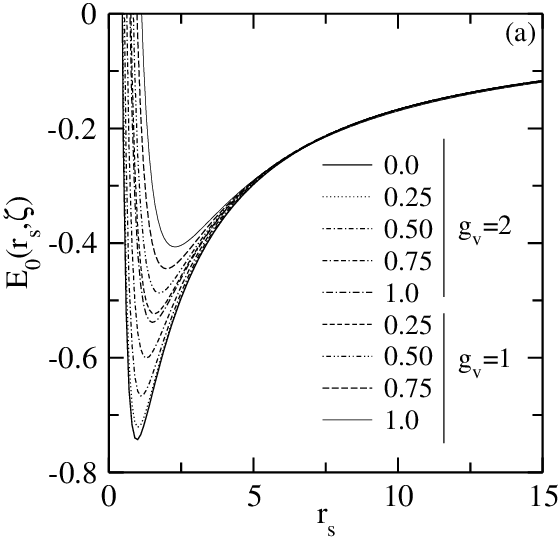}\vspace{5mm}
\includegraphics[width=65mm,height=60mm]{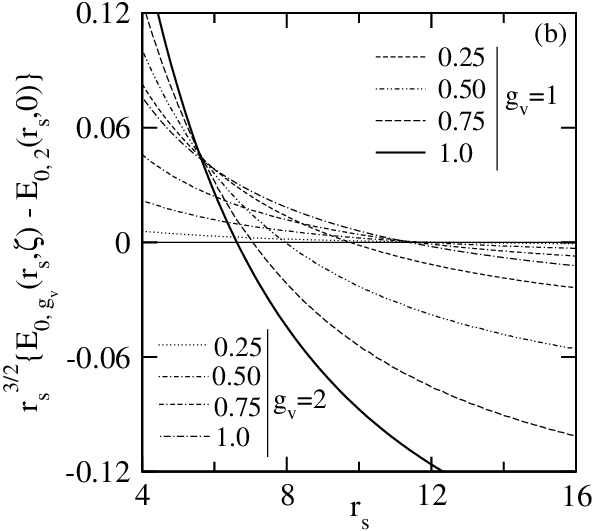} \caption{ \label{fig7}
The curves are labeled in the same manner as in \mbox{Fig. \ref{fig6}} except
that  the results are for the ideal 2DES.} 
\end{figure}
It is important to mention here that we have computed the potential energy by
performing  integration over $\lambda$ and not over the  $r_s$
parameter. This is rather necessary for the quasi 2DES
as the form factor $F(q)$ here is a function of $r_s$.  
Further, since the calculation of $E_0$ involves only $S_{CC}(q)$ and the STLS
theory yields a reliable $S_{CC}(q)$ despite the emergence of 
unphysical poles
in $\chi_{\sigma\sigma'}(q,\iota\omega)$ for $r_s$ above $r_s^c$, we
trust that our results of 
$E_0$ should be reliable for  $r_s > r^c_s$ also. 
\begin{figure}[t!]
\includegraphics[width=65mm,height=60mm]{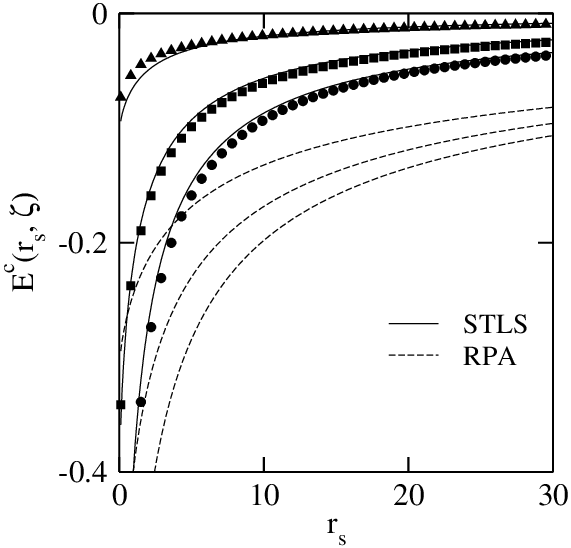}\vspace{5mm}
\caption{\label{fig8} The correlation energy $E^c(r_s,\zeta)$ vs $r_s$ for an
 ideal 2V- and 1V-2DES (from bottom to top) at $\zeta=0$ and $1$  
within the STLS and RPA 
approximations. The symbols represent the QMC data
taken from the Refs. \onlinecite{attacc} (1V) and \onlinecite{marchi} (2V).}
\end{figure}
We have checked this point explicitly at $\zeta=0$ by comparing $E_0$
computed  via $S_{\sigma\sigma'}(q)$ and through the direct calculation of
$S_{CC}(q)$ which does not involve poles and is possible only at
$\zeta=0$ and $1$. To our satisfaction, the two results matched within
the tolerance of self-consistent calculation. 
\par
In Fig. \ref{fig5}(a) we report $E_0(r_s,\zeta)$ for the quasi 2V2DES at
selected $\zeta$ over a wide range of $r_s$. Owing to overlapping
of different energy curves for 
$r_s\gtrsim 7$, we have depicted in Fig. \ref{fig5}(b) 
the energy difference $\{E_0(r_s,\zeta)-E_0(r_s,0)\}$ at different
$\zeta$ to resolve the stable spin phase. 
One may note that the unpolarized ($\zeta=0$) phase remains
stable up to  $r_s \approx 13.2$ 
and then, there occurs an  abrupt transition to
the polarized ($\zeta=1$) phase, with the partially spin-polarized
states remaining unstable. A qualitatively similar result is
found to hold for the quasi 1V2DES (see Fig. \ref{fig6}), where the
spin-polarization transition takes place at $r_s\approx 8$.
However, in order to get a clear picture of the ground state, we
must analyze the 1V and 2V cases together. Accordingly,
we plot in Fig. \ref{fig6} the energy difference 
$\{E_{0,g_v}(r_s,\zeta)-E_{0,2}(r_s,0)\}$  for $g_v=2$ and $1$ at
different $\zeta$.  
It may be pointed out that $E_0$ for the polarized 2V state is
equal to that for the unpolarized 1V state.
Apparently, the unpolarized 2V phase represents the ground state for
$r_s<8.8$, and there occurs at $r_s\approx 8.8$ a first-order
transition to the polarized  1V phase (i.e., simultaneous
valley- and spin-polarization) before there
could happen a transition to the polarized 2V  phase. Thus, we find that
the partially spin-polarized
phases are unstable in the Si (100) inversion layer.  
Moreover, our prediction of the unpolarized
2V ground state for $r_s < 8.8$ is in agreement with the 
experiment of Pudalov {\it et al.} \cite{puda1},
where the electron states have been observed to be four-fold degenerate 
over the investigated density range of $1.5<r_s<8.4$. 
\par
\begin{figure}[t!]
\includegraphics[width=65mm,height=60mm]{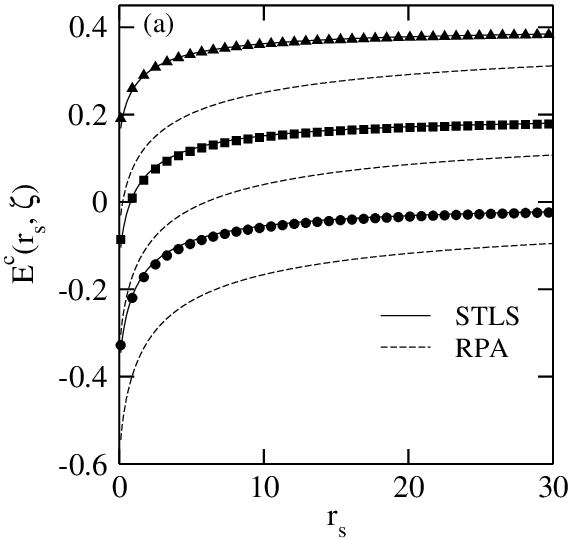} 
\includegraphics[width=65mm,height=60mm]{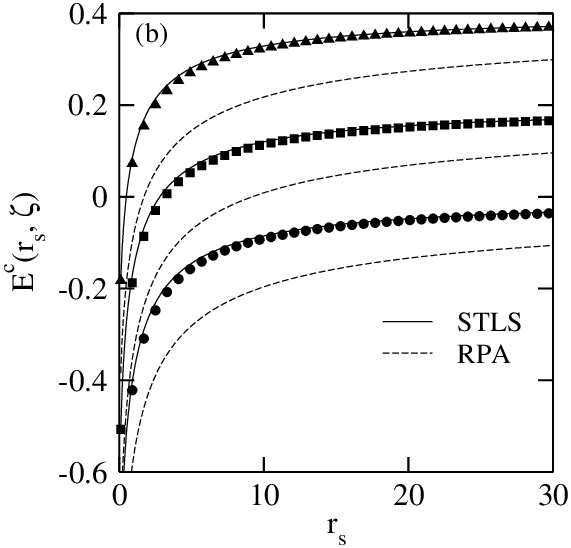} 
\caption{\label{fig9}
The correlation energy $E^c(r_s, \zeta)$ vs $r_s$ for an ideal 2DES at  
$\zeta$=0.25, 0.50, and 0.75 (from bottom to top) within the STLS 
 and RPA approximations for $1V$ [in panel (a)] and
 $2V$ [in panel (b)]. The symbols represent 
the QMC data taken from the Refs. \onlinecite{attacc} (1V) and 
Ref. \onlinecite{marchi} (2V). For clarity, $E^c(r_s,\zeta)$ at 
$\zeta$=0.50 and 0.75 has been shifted vertically 
by 0.2 and 0.4, respectively.} 
\end{figure}

For assessing the role of the finite width of 2DES, 
we report in  Fig. \ref{fig7} $E_0(r_s,\zeta)$ for the ideal 2DES.
Qualitatively, a phase diagram similar to the quasi
2DES is obtained, with the simultaneous valley- and spin-polarization
now occurring at $r_s \sim 6.7$.
In agreement with the QMC study \cite{attacc,marchi}, the partially
spin-polarized states are unstable irrespective of the valley degeneracy. 
Analyzing separately the 1V and 2V results,
we notice however that both support a first-order spin-polarization 
transition
at a sufficiently low $r_s$. 
In the 1V2DES this transition agrees
qualitatively with the QMC study \cite{attacc}, though the $r_s$
for the transition is underestimated by about a factor of $4$; however its occurrence 
in the 2V2DES
grossly disagrees with the QMC
simulations \cite{marchi,conti} where the unpolarized 2V
phase has been found to represent the ground state  for densities down
to the Wigner crystallization. To understand this qualitative 
mismatch, we compare in Figs. \ref{fig8} and \ref{fig9} the STLS
correlation energy $E^c(r_s,\zeta)$ with the  
QMC data \cite{marchi,attacc} at
different $\zeta$ and $g_v$. The RPA results are also shown to
demonstrate the importance of short-range correlations. 
Evidently, the STLS results are in very good agreement with
the QMC study both for $g_v=1$ and $2$, and  the RPA badly
overestimates the correlation energy. Thus, the situation remains unclear
from this comparison. 
\begin{figure}[h!]  
\includegraphics[width=65mm,height=60mm]{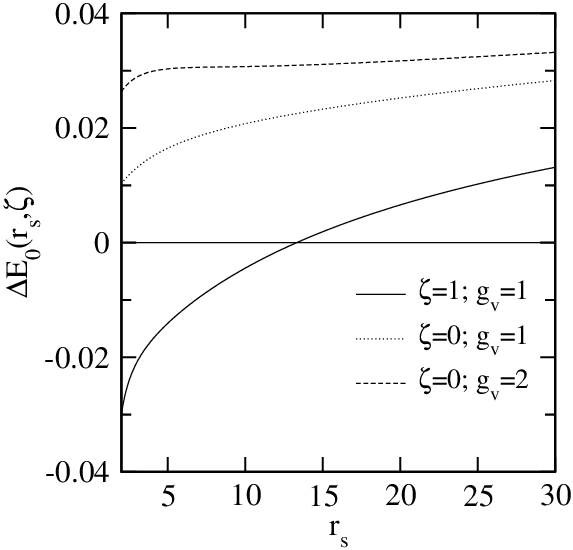}
\caption{\label{fig10} Fractional energy difference $\Delta E_0$ 
between the QMC and STLS
results of ground-state energy [i.e., 
$\{E_0^{QMC}(r_s,\zeta) - E_0^{STLS}(r_s,\zeta
)\}/E_0^{QMC}(r_s,\zeta)$]  vs $r_s$
at indicated $\zeta$ and $g_v$.  The QMC data
is taken from the Refs. \onlinecite{rapisarda} (1V) and
\onlinecite{conti} (2V).}  
\end{figure}
\par
To get further clue, we analyze in \mbox{Fig. \ref{fig10}} the
fractional difference
between the QMC and STLS ground-state energy  
$\Delta
E_0$=$\{E_0^{QMC}(r_s,\zeta)-E_0^{STLS}(r_s,\zeta)\}/E_0^{QMC}(r_s,\zeta)$
for the  1V and 2V systems at 
$\zeta= 0$ and $1$.  
Interestingly, the STLS theory underestimates the magnitude of 
$E_0$  for the unpolarized 1V and 2V states 
(with its extent increasing with $r_s$ and $g_v$), while for the polarized
1V state it exhibits overestimation in the $r_s$-region where the
STLS predicts the valley- and/or spin-polarization transitions. 
Apparently, it  is the
opposite sign of $\Delta E_0$ for the unpolarized  2V and the polarized
1V states which leads to simultaneous valley- and
spin-polarization at $r_s\approx 6.7$. On the other hand, the
spin-polarization in the 2V2DES seems to arise due to 
a decrease in  $\Delta E_0$ in going from $\zeta=0$ to $1$.
\subsection{Spin-resolved interaction energy of correlation}
In order to test further the STLS $S_{\sigma\sigma'}(q)$, we
compute and compare with the available QMC data (as fitted analytically by
Gori-Giorgi {\it et al.}\cite{pao}) the correlation contribution to the
spin-resolved interaction  energy $v^c_{\sigma\sigma'}(r_s,\zeta)$, for
an ideal 1V2DES.  
By definition
\begin{equation}
v^c_{\sigma\sigma'}(r_s,\zeta)=
E_{\sigma\sigma'}^{int}(\lambda=e^2;r_s,\zeta)-
E_{\sigma\sigma'}^{x}(r_s,\zeta ),
\end{equation}
 with  
\begin{equation}
E_{\sigma\sigma'}^{x}(r_s,\zeta)=-\delta_{\sigma\sigma'}\frac{4\sqrt{2}}{3\pi
  r_s}\left[1+(-1)^{\delta_{\sigma\downarrow}}\zeta \right]^{3/2},
\end{equation}
\begin{figure}[b!]
\includegraphics[width=55mm,height=50mm]{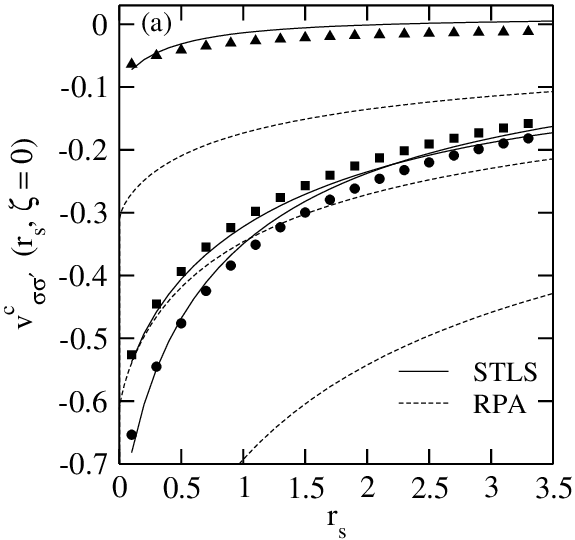}\\
\includegraphics[width=55mm,height=50mm]{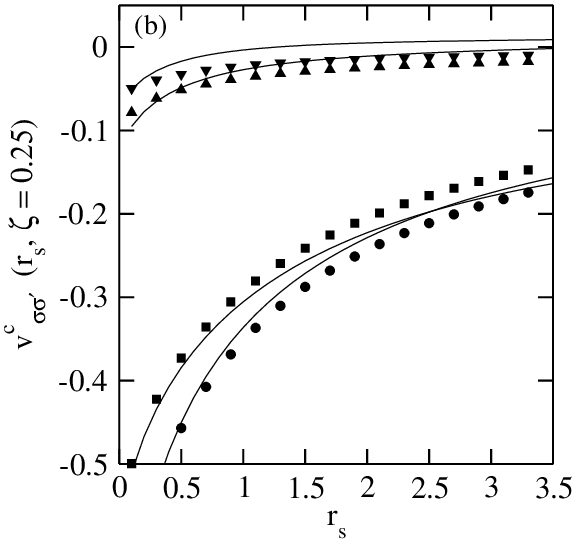}\\
\includegraphics[width=55mm,height=50mm]{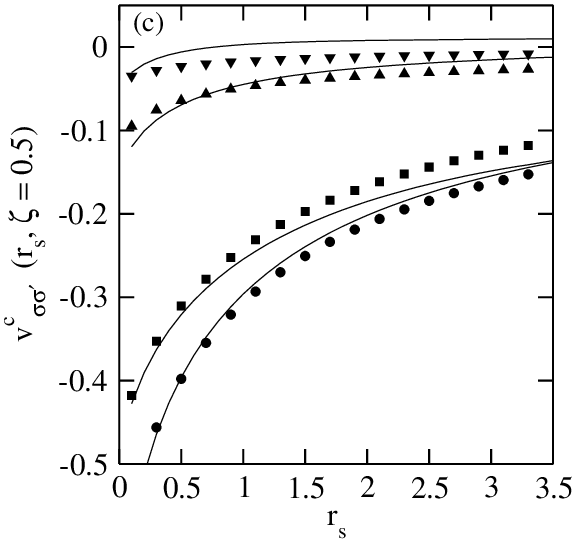}\\
\includegraphics[width=55mm,height=50mm]{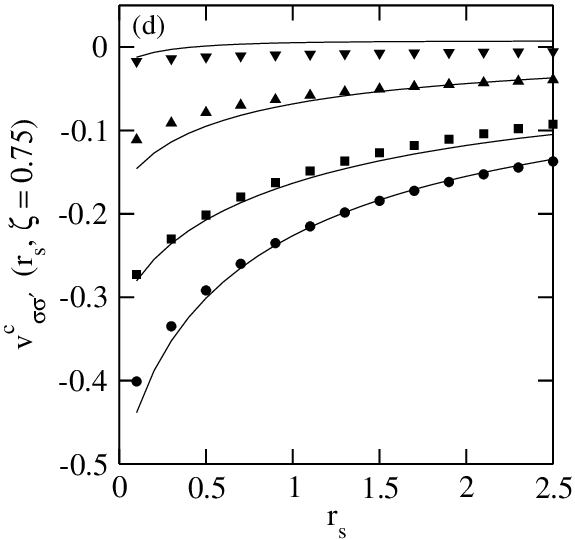}
\caption{\label{fig11} The spin-resolved interaction energy of 
correlation $v^{c}_{\sigma \sigma'}(r_s, \zeta)$ vs $r_s$ at selected 
$\zeta$ for an ideal 1V2DES. The symbols are the QMC data as fitted by 
Gori-Giorgi {\it et al.}  \cite{pao} In each case, the curves and
symbols from top to bottom represent, respectively, 
the $\downarrow\downarrow$, 
$\uparrow\uparrow$, $\uparrow\downarrow$, and the spin summed
(i.e., $v^c_{\uparrow\uparrow}+v^c_{\downarrow\downarrow}+
v^c_{\uparrow\downarrow}$) contributions.} 
\end{figure}
being the exchange contribution to the interaction energy.
$v^c_{\sigma\sigma'}(r_s,\zeta)$ are plotted in 
\mbox{Fig. \ref{fig11}}  as a function of $r_s$ at different $\zeta$.
Results are shown for $r_s$ up to $r_s^c$, because thereafter
$S_{\sigma\sigma'}(q)$ are not reliable due to unphysical poles in 
$\chi_{\sigma\sigma'}(q,\iota\omega)$. Notably, the STLS theory
underestimates the $\uparrow\uparrow$ and 
$\downarrow\downarrow$ contributions, while the $\uparrow\downarrow$
part is overestimated to an extent that the spin summed
result (i.e., $v^c_{\uparrow\uparrow}+v^c_{\downarrow\downarrow}+
v^c_{\uparrow\downarrow}$) lies close to the QMC data.
The degree of underestimate is more for the $\downarrow\downarrow$
(i.e., the dilute spin) component, with its value becoming small
positive against the QMC result beyond a certain $r_s$.  
The cross-spin correlation dominates over the like-spin contribution
meaning thereby the latter are  determined mainly by the exchange
effects. Furthermore, a comparison with the RPA reveals that 
it overestimates all three correlation components, with the like-spin 
part showing the maximum departure from the QMC data; this point is
illustrated in Fig. \ref{fig11}(a) at $\zeta=0$. 

\begin{figure}[b!]  
\includegraphics[width=65mm,height=60mm]{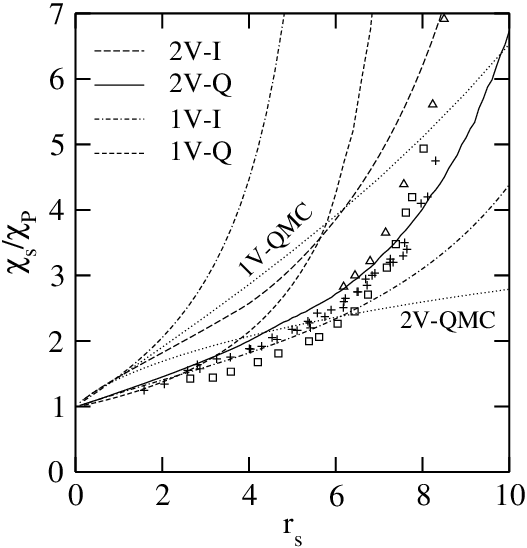}
\caption{\label{fig12}  The spin susceptibility enhancement $\chi_s/\chi_P$ vs
$r_s$   for the quasi (Q) and ideal (I) models of the Si (100)
inversion layer. The symbols represent the experimental 
data of Pudalov {\it et al.} \cite{puda2} and Shashkin {\it et al.} 
 \cite{shas2} The dotted curves represent,
 respectively, the QMC results for the ideal 1V
 (Ref. \onlinecite{attacc}) and 2V (Ref. \onlinecite{marchi}) 2DESs. 
The dashed-dot curve is the RPA result for the quasi Si (100) inversion
layer.}  
\end{figure}

\subsection{Spin susceptibility} 
The spin susceptibility $\chi_s$ is computed (from Eq. (\ref{e15})) 
for three different
devices, viz. the Si (100) inversion layer, the AlAs
QW, and the GaAs HIGFET. 
For the AlAs QW, we employ the infinite QW model where\cite{gold2} 
\begin{equation} F(q)= \frac{1}{4\pi^2 +q^2a^2} \left (3qa + 
\frac{8 \pi^2}{qa} - \frac{32 \pi^4}{{q^2a}^2} \frac{1- e^{-qa}}
{4\pi^2 +{q^2a}^2}\right ),
\end{equation}
with $a=4.5nm$ being the width of the well. In case of the GaAs HIGFET,
$F(q)$ can be obtained \cite{zhang2} from Eq. (\ref{e5}) by setting 
$\epsilon_{ins}=\epsilon_{sc}$ and  
taking $\epsilon=12.9$, $m_z=0.067m_e$, and $n_d=0$\cite{depalo}.  
\par
Figure \ref{fig12} depicts our $\chi_s/\chi_P$ for the Si (100) 
inversion layer, along with the experimental data of 
Pudalov {\it et al.} \cite{puda2} and Shashkin {\it et al.}
\cite{shas2}, and the QMC and STLS results for an
ideal 2V2DES model. To see the effect of valley degeneracy, the STLS and
QMC results \cite{attacc}  are also given at $g_v=1$. 
However, we should look for a comparison at $g_v=2$ as the 2V
unpolarized phase is found to represent the ground state over the
$r_s$ range where experimental $\chi_s$ is available. 
It is gratifying to note that $\chi_s$ for $g_v=2$ exhibits very
good agreement with the experiment, except at the largest $r_s$ values, near the
apparent ferromagnetic instability found in  Ref. \onlinecite{shas2}. Importantly,
it is crucial to
include in the theory both the valley degeneracy and the finite
thickness of the 2DES. Both of these factors act to suppress
$\chi_s/\chi_P$ appreciably over the ideal 1V2DES estimate. Indeed, considering
the sizeable disagreement between STLS and QMC predictions for the strictly
2D systems, the good agreement  found with the experiment up to $r_s \approx 7.5$
can only be understood in terms of the dominance of  thickness over correlation
effects\cite{depalo}. Similar results
have been reported recently by De Palo {\it et al.} \cite{depalo} and
Zhang and  Das Sarma \cite{zhang}. 
\par
Although not reported here, it is consistent to find that $\chi_s$ tends
to diverge near the spin-polarization
transition. For the Si (100) system, the divergence occurs at
$r_s\sim 13$, which lies close to 
an experimental estimate\cite{pudav2} of the  
critical $r_s$ for the ferromagnetic instability.    
\begin{figure}[t!]
\includegraphics[width=65mm,height=60mm]{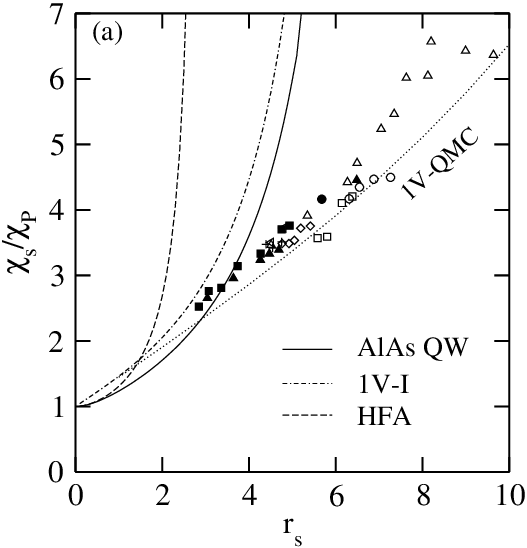}
\includegraphics[width=65mm,height=60mm]{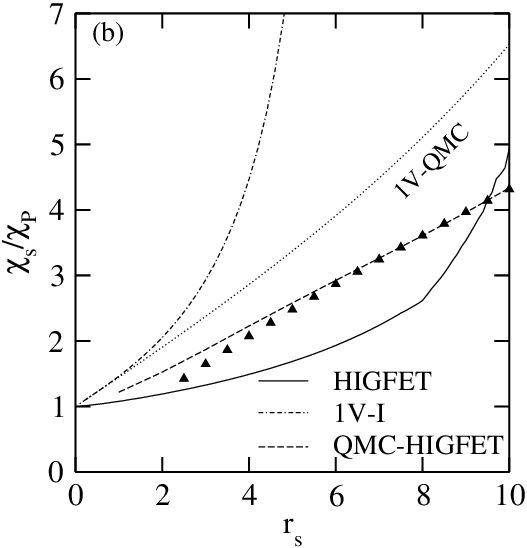} 
\caption{\label{fig13} The spin susceptibility enhancement $\chi_s/\chi_P$ vs $r_s$
for the AlAs QW [in panel (a)] and the GaAs HIGFET [in panel (b)],
along with the STLS (double dashed-dot lines) and QMC
\cite{attacc} (dotted lines) 
results for an ideal (I) 1V2DES. 
The symbols represent the respective experimental results of
Vakili {\it et al.} \cite{vakili} and Zhu {\it et al.} \cite{zhu}
In (a) dashed curve is the HFA result for an ideal 1V2DES, 
while in
(b) it refers to the thickness corrected QMC result for the GaAs
HIGFET due to Depalo {\it et al.} \cite{depalo} } 
\end{figure}
\par
Next we show in Fig. \ref{fig13} $\chi_s/\chi_P$ for the AlAs QW and
the GaAs HIGFET, together with the respective
experimental data \cite{vakili,zhu} and the STLS results for
 the ideal 1V2DES. 
Once
again we note that the inclusion of the finite width of the 2DES
brings the STLS predictions closer to the experiment.
However, the AlAs QW being much narrower exhibits comparatively
smaller suppression due to finite thickness and  the agreement is much 
less satisfactory than for the Si (100) system, 
both for the AlAs QW and the GaAs HIGFET. 
\par
Interestingly, the quality of the STLS
predictions is different in three systems. This seems to suggest that
it might be important to include 
other device specific parameters like disorder  
\cite{waintal}.
Moreover, the choice of confining potential is
expected to affect the form factor and hence, the interaction
potential; for instance, taking a finite (AlAs) QW model would result in
a softer potential and therefore, in suppression of $\chi_s$ over the
infinite QW result. Likewise, in the inversion layer model 
Stern and Howard \cite{stern} did not
consider the exchange-correlation effects in
calculating the  interface electronic eigenfunctions.  Later on Ando
\cite{ando2} found that these effects reduce the width  of
inversion layer, which in turn is expected to enhance $\chi_s$.  
\section{Summary and conclusions}
To summarize, we have investigated the
spin properties of a 2DES, as realized at
the interface of semiconductor-insulator/semiconductor heterojunctions,
by using the STLS approach. 
The Si (100) inversion layer has been studied in detail to 
calculate the spin-resolved static structure factors and the ground-state
energy at arbitrary $\zeta$ over a wide range of $r_s$. 
We deduce quite generally that the STLS theory fails in determining the
spin-resolved correlations above a critical $r_s$ due to 
the appearance
of unphysical poles in $\chi_{\sigma\sigma'}(q,\iota\omega)$,
i.e., at imaginary frequency. Such poles, however,  may be shown to not
contribute to the frequency integral of $\chi''_{\sigma\sigma'}(q,\omega)$
yielding the structure factors, which can still be computed,  and the spin
summed properties  (such as the charge-charge structure factor) remain 
reliable and close to the available QMC simulation data. 
\par
This enabled us to determine the
ground-state energy over a wide range of $r_s$ at arbitrary
$\zeta$ and hence, the stable spin phase. For the 
quasi Si (100) system,
the unpolarized 2V phase is found to represent the ground state for
$r_s\lesssim 8.8$ and thereafter, a first-order transition occurs to
the polarized 1V phase.  
The predicted  ground state for $r_s\lesssim 8.8$ agrees with 
the  experiment of Pudalov {\it et  al.}\cite{puda1}
Furthermore, the calculated spin susceptibility agrees fairly well with the
experiment, but only with the inclusion of transverse thickness and  
valley degeneracy, provided one does not come too close to  the
apparent ferromagnetic instability found in  Ref. \onlinecite{shas2}.
\par
Without thickness we find that STLS sizeably overestimates the spin
susceptibility, both for the 1V and 2V states, a drawback in common with 
RPA\cite{depalo,marchi}.
The spin susceptibility has also been calculated for the 2DESs as realized 
in the AlAs QW and the GaAs HIGFET. We find that   agreement with experiment
is much less satisfactory in these cases, even though it is somewhat improved 
by the inclusion  of thickness. 
\par For the ideal (i.e., zero thickness) 2DES model, the
STLS theory  reproduces nicely the recent QMC correlation energy at
arbitrary $\zeta$ both for the 1V and 2V states. 
However, the agreement is seen to depend qualitatively on $\zeta$ and
$g_v$. Particularly, the correlation energy is
overestimated for the 1V state at $\zeta=1$, whereas it is underestimated
for the 2V state both at $\zeta=0$ and at $ \zeta =1$. 
Although the magnitude of
deviation from the QMC data is small, yet this being of opposite sign
causes simultaneous valley- and
spin-polarization at $r_s \sim 6.7$, a prediction in gross
violation of the QMC study. An analysis of the
correlation contribution to the spin-resolved interaction energy
reveals that the STLS theory is
relatively less accurate in handling the like-spin correlations. 
This result may have bearing on the further development of the theory
of electron correlation.
\section*{ACKNOWLEDGEMENTS}
The authors are grateful to Paola Gori-Giorgi for providing the QMC
simulation data of
spin-resolved pair-correlation functions.
\end{document}